%Paper: hep-th/9209132
%From: "Judy Mack, University of Rochester, 716-275-4840"
% <JUDY%UORHEP.bitnet@CUNYVM.CUNY.EDU>
%Date: Wed, 30 Sep 1992 12:03 EST

\magnification=1200
\baselineskip=20pt
% 20 pt is double spacing (18 pt is 1.5 spacing).
\def\lsim{<\kern-2.5ex\lower0.85ex\hbox{$\sim$}\ }
\def\rsim{>\kern-2.5ex\lower0.85ex\hbox{\sim$}\ }
\overfullrule=0pt
\ \ \ \
\vskip 2cm

\centerline{\bf Comment on \lq\lq Internal Frame Dragging and a Global}
\centerline{\bf Analog of the Aharonov-Bohm Effect"}
\vskip 2cm
\centerline{by}
\vskip 2cm
\centerline{C. R. Hagen}
\centerline{Department of Physics and Astronomy}
\centerline{University of Rochester}
\centerline{Rochester, NY 14627}
\vfil\eject
In a recent work$^1$ it has been claimed that Aharonov-Bohm (AB)
scattering$^2$ can occur for sufficiently low momentum transfers in the
presence of a symmetry breaking vortex configuration.  According to the
argument given there crucial elements for the occurrence of this effect
are the existence of two degrees of freedom as well as an angular dependence
(\lq\lq internal frame dragging") of the symmetry breaking term.  This note
will point out, however, that neither of these is essential to obtaining an
AB form for the scattering cross section in the forward direction and that
the effect is rather a consequence of the extreme long range behavior of the
symmetry breaking term.  In the examples considered$^1$ -- both in the
scalar and spinor cases -- infinite energy configurations are
required to generate the AB result.

The symmetry breaking effect in ref. 1 invokes a scalar field, which is
taken to have expectation value $ve^{i \phi}$ outside a core region.  Since
the $\lambda$ field contributes an amount $\vert \partial \lambda \vert^2$
to the energy of the system, it is straightforward to obtain the result that
at large $r$ it gives a logarithmically divergent contribution to the
energy.  If one elects to cut off the large $r$ behavior, this
divergence can be avoided, but the $(1/\sin^2 \theta/2)$ form of the cross
section in the forward direction is necessarily lost.$^3$  The result is
similar in the spin one-half case.  The constant magnitude of the
magnetization which is essential to the result is well known$^4$ to imply
logarithmically divergent self energy for an isolated vortex.

On the other hand it is easy to show in specific (finite energy) examples
that there exist cases in which neither $\phi$ dependent symmetry breaking
nor multiple degrees of freedom are required to generate the AB cross
section.  Consider, for example, the model in which $< \lambda>\> \sim \>
{\rm const}/r$ for large $r$.  This choice implies a finite
ground state energy for the $\lambda$ field, but at the same time provides
an effective potential which has sufficiently long range to imply a
divergent forward cross section.  This result has in fact been
 calculated$^5$ and
 is known to yield precisely the $(1/\sin^2 \theta/2)$ AB form  for
small scattering angles.

One notable difference between this work and ref. 1 is that in the latter
case \lq\lq an essentially geometrical, parameter-independent form" is
obtained for the cross section whereas the $\phi$ independent symmetry
breaking discussed here implies that it should be proportional to the
square of the symmetry breaking parameter.  Since, however, physical
quantities are frequently found to become parameter independent when
certain unphysical limits are taken, it is reasonable to conclude that
the parameter independence of the results of ref. 1 is principally a
consequence of the infinite energy of the system considered there.  In
fact it seems that $\phi$ independent symmetry breaking leads in a much
more natural way to AB-like cross sections.  In particular, a $\phi$
dependence appears to inhibit the effect, forcing one to postulate
considerably more singular configurations in order to generate the same
phenomenon.
\medskip
\noindent {\bf Acknowledgments}

The author is indebted to S. Teitel for conversations on the XY model and
to the D.O.E., Grant No. DE-FG02-91ER40685.

\bigskip
\noindent {\bf References}
\item{1.} J. March-Russell, J. Preskill, and F.  Wilczek, Phys. Rev. Lett.
{\bf 68}, 2567 (1992).
\item{2.} Y. Aharonov and D. Bohm, Phys. Rev. {\bf 115}, 485 (1959).
\item{3.} When such a cutoff is used, it may be possible to obtain an AB
form of the cross section, if only at \underbar{intermediate} momentun
transfers.  Even this would require that there exist a nonvanishing range
of momentum transfers which are much less than the $\Gamma$ of ref. 1 but
much greater than the square of the inverse cutoff distance.  Such a range
may or may not exist, depending upon the particular application.
\item{4.} M. Plischke and B. Bergensen, \underbar{Equilibrium Statistical
Physics},  Englewood Cliffs, N.J.: Prentice Hall (1989).
\item{5.} J. Law, M. K. Srivastava, R. K. Bhaduri, and A. Khare, J. Phys.
 A {\bf 25}, L183 (1992).
\bye